\documentclass{article}
\usepackage[dvips]{graphicx} 
\usepackage{amssymb} 

\def\kT{\kappa T}

\begin{document}

\begin{center}
{\large Stochastic dynamics of magnetic nanoparticles and a mechanism of biological orientation in the geomagnetic field}\\~ \\
V.N. Binhi%
\footnote{Correspondence:  GPI RAS, 38, Vavilova St., Moscow, 119991, Russia.  Tel./Fax: +7(095)216-4819; e-mail: vnb@biomag.info} 
\\ ~\\
\em{A.M. Prokhorov General Physics Institute RAS, Moscow}
\end{center}

\[
\begin{minipage}{11cm}
\noindent\small
The rotations of microscopic magnetic particles, magnetosomes, embedded into the cytoskeleton are considered. A great number of magnetosomes are shown to possess two stable equilibrium positions, between which there occur transitions under the influence of thermal disturbances. The random rotations attain the value of order of radian. The rate of the transitions and the probability of magnetosomes to stay in the different states depend on magnetic field direction with respect to an averaged magnetosome's orientation. This effect explains the ability of migrant birds to faultless orientation in long-term passages in the absence of the direct visibility of optical reference points. The sensitivity to deviation from an `ideal' orientation is estimated to be 1--2 degrees. Possible participation of magnetosomes in biological effects caused by microwave electromagnetic fields is discussed.
\vspace{5mm}

\noindent {\em magnetosome, biogenig magnetite, magnetic compass orientation, migrant birds, magnetic field}
\end{minipage}
\]
\vspace{10mm}

Many birds of passage and other migrant animals overcome annually thousands of miles and accurately find the places of their seasonal habitats. This fact is not completely understood yet. A number of hypothesis have been suggested to account for this phenomenon \cite{kirschvink85-e}. In particular, the optical orientation and navigation based on `maps' of the terrestrial surface and starry sky are studied. Also studied is the navigation along the lines of force of the geomagnetic field, which are known to be rigidly bound with the geophysical coordinates of the Earth. The latter hypothesis finds the experimental validation in the numerous facts of the ability of some biological species among microorganisms, insects, fishes, birds, and mammals to orient in a magnetic field (MF) or react to the change in its direction with respect to other acceptable reference marks. 

At the same time, there is no recognized explanation for this phenomenon yet. Magnetic orientation is a part of the more general problem of the biological efficacy of weak, less than 1 G, magnetic fields (MFs). A brief review of the theoretical works in this area may be found in \cite{binhi03ae}, and the detailed discussion in \cite{binhi02}. 

The problem is in that magnetic energy of biologically active molecules in the geomagnetic field is very small. It does not exceed the energy of the electron magnetic moment in the Earth magnetic field $0.5\times 10^{-20}$~erg. This is more than seven order less than that of thermal fluctuations, i.e. $\kT \approx 4.1 \times 10^{-14}$~erg at physiological temperatures. It is not clear, how could such a small 'signal' cause a biological reaction on the thermal 'noise' background. 

However, there are submicron particles, which have magnetic moments. They have been found in many living objects, particularly in those displaying magnetic navigation ability. They consist of the magnetite mainly. Magnetic moment $\mu $ of the particles exceeds the elementary one in 7--9 orders. The energy of their turn in a weak magnetic field $H$ is essentially larger than that of thermal fluctuations. For single-domain magnetite particles of radius $r = 10^{-5}$~cm or 100~nm in the geomagnetic field the energy $\mu H \approx vJ H$ equals approximately $24 \kT $, where $\mu$ is the magnetic moment of the particle, $v$ and $J\approx 480$~G are the volume and the saturation magnetization. 

The cytoplasm near cell membranes is such that the turning of a microparticle may serve as a stimulus to cell division or ignite a nerve impulse. For example, MF produced by a particle with the afore-cited parameters is of value up to 0.1~T near the particle itself and strongly depends on its orientation. Therefore, the turn of particle may appreciably change the rate of some chemical reactions with the participation of free-radical pairs. 

Particularly interesting are the magnetite particles found in the brain of many animals and in human brain. The nerve tissue of the brain is separated from the circulatory system by the blood-brain barrier which is impermeable for most chemicals. In turn, the circulatory system is separated from the digestive system. Therefore, relatively large ferro- or ferrimagnetic particles cannot penetrate into brain tissue as a pollutant. They are found to have a biogenic origin, i.e. they appear over time as a direct result of the crystallization in brain matter. Biogenic magnetite particles are often called `magnetosomes'; they were first discovered in bacteria that displayed magnetotaxis \cite{blakemore75}. It was recently shown that magnetic nanoparticles may be produced in DNA complexes \cite{khomutov04e}. Biogenic magnetite undoubtedly plays role in the navigation of migrant birds \cite{kirschvink85-e}, insects \cite{etheredge99}, and in other cases.

For the explanation of the magnetic navigation, the dynamics of magnetosomes was modelled by using the equation of free rotations in a viscous liquid. This regimen was assumed to be most favorable from the viewpoint of the magnitude of possible effects. However, the assumption about free rotations does not match the data of electronic microscopy. For example, in some microorganisms, magnetosomes are assembled into firm univariate chains, where rotations are impossible. The idea was used as well that rigidly bound magnetosome brings pressure on a closely set receptor. 
However, in this case, the energy is transferred to the number of molecules simultaneously, so that just a small amount of energy relates to each one as compared to the $\kT $. 

It turns out that taking into account the not-too-strict elasticity of the medium enables one to describe qualitatively new stochastic rotational dynamics of magnetosomes that is useful to explain the magnetic navigation.

This article considers the dynamics of a magnetite particle embedded in the cytoskeleton. The latter consists of a 3D net of protein fibers of 6 to 25~nm in diameter that include actin filaments, intermediate filaments, and microtubules. The ends of these fibers may be fastened to the membrane surface and to various cell organelles. We assume the fibers may also be fastened to a magnetosome surface normally covered with a bilayer lipid membrane \cite{gorby88}. This fixes the position of the magnetosome and constrains its rotation to some extent. The stationary orientation of the magnetosome generally does not follow the constant MF direction. The balance of the elastic and `magnetic' torques determines the orientation now. The torque $\bf m$ affecting a particle of the magnetic moment $\boldmath \mu$ in an MF $\bf H$ equals ${\bf m} = \boldmath \mu \times {\bf H}$.   

\begin{figure}[t]
\[ \includegraphics[width=0.7\textwidth]{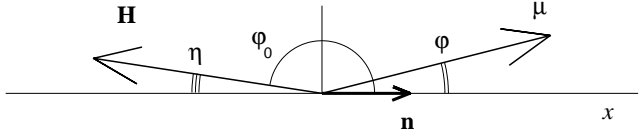} \]
\caption{\label{f01} {~~} {\protect \parbox[t]{100mm}{Relative arrangements of the vectors of magnetic fields and the magnetic moment of a magnetosome.} }}
\end{figure}

Here, putting aside the 3D character of the magnetosome rotations, we consider the magnetosome's motion in the plane of two vectors: the unit vector $\bf n$ of the $x$-axis, with which the vector of magnetosome's magnetic moment coincides in the absence of the MF (equilibrium position, $\varphi =0$), and the MF vector $\bf H$, Fig.~\ref{f01}. 

The Langevin equation for rotational oscillations of the particle is as follows: 
\begin{equation} \label{01} I \ddot{\varphi } + \gamma \dot{\varphi } + k \varphi = -\mu H(t) \sin(\varphi - \varphi_0) + \xi'(t) ~,~~~ \omega_0=\sqrt{k/I}~, \end{equation} where $\varphi$ is the angular displacement, $I$ is the moment of the particle's inertia, $\gamma$ is the dissipation coefficient, $k$ is the factor of mechanical elasticity resulting from the cytoskeleton fibers' bending, $\xi'(t)$ is a stochastic torque with the correlation function $\langle \xi'(t) \xi'(t+\Delta t) \rangle = 2\gamma \kT  \delta(\Delta t)$, while $\omega_0$ is the eigenfrequency, and $\varphi_0$ is the MF direction.
Then, we assume the quantity of fibers fastening the magnetosome to the cytoskeleton may vary from particle to particle and a significant number of magnetosomes are mobile enough to markedly change their orientation in the geomagnetic field. This means the mechanical elasticity due to the fibers' bending is of the same order as or less than the magnetic elasticity $k \lesssim \mu H \approx 24 \kT $. For magnetite Fe$_3$O$_4$ particles with the substance density $\rho \approx 5.2$~g/cm$^3$ and radius $r \approx 10^{-5}$~cm, we derive a value $\omega_0$ in the order of $10^{6}$~rad/s. A resonance, however, is not possible since the inertia forces are much less than viscous forces: $ I\omega_0 \ll \gamma $. Hereafter, the inertia term in the equation of motion may be ignored.

The idea of this work is to study the dynamics of a magnetosome fixed into a visco-elastic cytoskeleton and predominantly oriented in a direction opposite to that of a constant MF. We assume further $\varphi_0 = \pi - \eta$. The angle $\eta $ should be read as the off-course angle, i.e. the azimuth deviation from an `ideal' or reference direction, which is determined by the mechanical bonds that fasten the magnetosome in an averaged position with regard, for example, to animal's cranium. In other words, in the frame of reference of the geomagnetic field, $\eta $ is the deviation of animal's orientation from an `ideal' one in parallel to the geomagnetic field vector.

\begin{figure}[t]
\[ \includegraphics[width=0.7\textwidth]{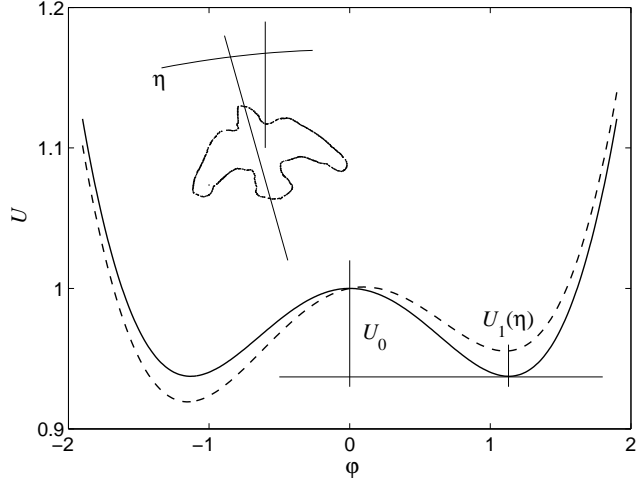} \]
\caption{\label{f02} {~~} {\protect \parbox[t]{100mm}{The potential function of a magnetosome at the off-course angle $\eta =0$ and $\eta \neq 0$. $U_0$ is the potential barrier height, $U_1$ is the change of the barrier height with an off-course deviation; the elasticity parameter equals $a=0.8$.} }}
\end{figure}

For small angles $\eta $ the equation of motion takes the form:
$ \gamma \dot{\varphi}  +k\varphi  = \mu H \sin(\varphi) + \mu H \eta \cos (\varphi) + \xi'(t)~. $ 
With the designations
\begin{equation} \label{99} a= \frac{k} {\mu H }~,~~ \tau \equiv \frac {\mu H} {\gamma} t~,~~ D\equiv \frac {2\kT} {\mu H} \end{equation} the equation is reduced to 
\begin{equation} \label{100} \dot{\varphi} + \partial_{\varphi} U(\varphi, \tau) = \sqrt{D} \xi(\tau) \end{equation} with the potential 
\begin{equation} \label{1001}
U(\varphi, \eta) = \cos (\varphi) + \frac a2 \varphi^2 - \eta \sin (\varphi ) ~ . \end{equation} Here $\xi(\tau)$ is the centered Gaussian process of unit variance (the identity $\delta(\alpha t)= \delta(t)/|\alpha | $ is used). 

The potential energy of a magnetosome in terms of $\mu H$ is shown in Fig.~\ref{f02}. As is seen, for not too large angles at $a<1$ there are two stable equilibrium positions $\varphi_{\pm}$ and the unstable one $\varphi_0 =0$. Due to thermal disturbances, there are transitions from well to well, the random turns being of significant value around 2~rad. 

\begin{figure}[t]
\begin{center}
\includegraphics[width=0.6\textwidth]{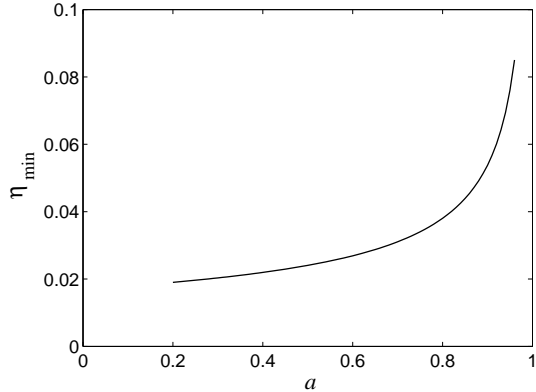} 
\end{center}
\caption{\label{f03} {~~} {\protect \parbox[t]{100mm}{The minimally detectable off-course angle depend upon the elasticity parameter.} }}
\end{figure}

Consider the joint influence on a magnetosome of a random torque $\xi(t)$ and a magnetic signal related with the deviation of the organism's orientation from the reference one. The magnetic signal, which varies directly with $\eta $, results in the change of the potential function, Fig.~\ref{f02}. The state of the magnetosome oriented in the direction of the absolute minimum of the potential function becomes a preferred one. The ratio of the probabilities of the magnetosome to locate in the states $\varphi_{\pm}$ equals
\begin{equation} \label{103} \frac{p_-} {p_+} =\exp \left( 2\frac {\delta U} {D} \right) ~, \end{equation}
where $\delta U= 2U_1$ is the potential difference of the equilibrium points. This implies: 
\[ p_- = \frac12 \exp \left( 2 \frac{U_1} {D}\right) ~. \]  The `signal' proportional to the difference of the probability and its equilibrium value 1/2 equals
\[ s \equiv p_- - \frac12 \approx \frac {U_1} D ~. \] Since the `noise' is the equilibrium value 1/2, the signal-to-noise ratio in this case is equal to 
\begin{equation} \label{031} R_{\rm sn} = 2 \frac {U_1} D ~. \end{equation}
The quantities $U_0$, $U_1$ of the potential (\ref{1001}) have no exact analytical presentation. Here we derive them as the expansions over the parameter $1-a$, which is assumed to be a small one:
\[ \varphi_{\pm }^2 = 6 (1-a),~~ U_0 = \frac 32 (1-a)^2 ,~~ U_1^2 = 6 \eta ^2 (1-a) ~,~~ U''(0)= a-1 ,~~ U''(\varphi_{\pm}) = 2(1-a) ~. \]
%
%
The minimally detectable angle of deviation from the reference course follows the equation 
\begin{equation}\label{032}
R_{\rm sn} =1 ~. \end{equation}
From this equation, substituting the derived value $U_1$ in (\ref{031}), we arrive to the formula:
\begin{equation} \label{104} \eta_{\rm min} = \frac D {2\sqrt{6(1-a)}} ~. \end{equation} This quantity is shown in Fig.~\ref{f03} as the function of the elasticity parameter $a$. As is seen, the maximum sensitivity takes place at small values $a$, i.e. for `softly' fastened magnetosomes. However, arbitrary small values of $a$ make no physical sense.

The expression (\ref{103}) and then (\ref{104}) are valid for the equilibrium probability distribution. This means the changes in the potential have to occur more slowly than the relaxation to a statistical equilibrium. One can separate relaxations within each potential well and between wells. For small values $a$, when the potential barrier is high, the relaxation time is determined mainly by the well-to-well transitions. The character time here is the mean first passage time (the Kramers time, see for example \cite{mcnamara89})
\[ \tau_{_{\rm K}} = \frac {2\pi} {\sqrt{|U'' (0)| U''(\pm \varphi) }} \exp \left( \frac {2 U_0} {D} \right) = \frac {\pi \sqrt{2}} {1-a} \exp \left[ \frac {3(1-a)^2} {D} \right] ~. \]
The equilibrium distribution takes place on condition that
\[\tau_{\rm or} = \frac {\mu H} {\gamma} t_{\rm or} \gg \tau_{_{\rm K}} ~, \] where $\tau_{\rm or}$ is a character time of the reorientation of animal, the periods of `hunting' around the reference course. Assuming $t_{\rm or} \ge 1$~s and taking $\gamma \approx 4\pi \nu r^3 \approx 20\times 10^{-17}$~erg$\cdot$s for the damping coefficient of the rotations of magnetosome with the radius $10^{-5}$~cm in a liquid with viscosity $\nu \approx 10^{-2}$~g/cm$\cdot$s (water), the condition is fulfilled if $a > 0.65$. Therefore, from Fig.~\ref{f03} is seen that the sensitivity to the course deviations is about 0.03~rad or 1.7 degree.

We will note, if the signal-to-noise ratio in (\ref{032}) was $n$ times greater, the minimally detectable angle would be $n$ times less.

To explain the observable sensitivity of animals to very weak geomagnetic variations, in \cite{yorke85-e} there was suggested that ensembles rather than single magnetoreceptors detect MF changes. Given that, the averaging of the signals coming from magnetoreceptors, which cause the signal-to-noise to enhance, occurs in the cerebrum. With averaging, the signals of regular rotations are summed; for chaotic rotations, what is summed are just the squares of the chaotic signals. Therefore, the enhancement of the signal-to-noise ratio is reached proportionate to the square root of the number of magnetosomes contributing to endpoint response.  We remind that the density of magnetosomes in human brain tissue was measured around $5\times 10^6$, and in brain meninges more than $10^8$ crystals per gram \cite{kirschvink92b}. Some birds have the concentration of the magnetite crystals in orders higher. 

There is another factor of the signal-to-noise enhancement that apparently was not reported in literature earlier. If we assume each single bird makes flight corrective turns during passage in accord with the flight path of the flock, i.e. of the most of birds, then the presence of an effective averaging mechanism follows that is taken out of the bird's organism. Evidently, this mechanism reduces the level of fluctuations in flight direction in proportion to the square root of the number of birds in flock.

Even without an additional enhancement of the signal-to-noise ratio, the retention of the orientation accurate to 1--2 degrees during passage could provide its exactitude since the additional course correction may occur due to the optical navigation in time periods of good visibility.

In conclusion, we note the following. (1) The stochastic dynamics of magnetosomes has peculiar features in ac MF (a stochastic resonance of magnetosomes) and in the slowly varying geomagnetic and `zero' magnetic fields as well. These features enable one to explain the biological effects of weak extremely-low-frequency MFs, geomagnetic storms, and a `magnetic vacuum'. (2) The Debye relaxation in water has a wide maximum in the microwave range of the electromagnetic radiation. The viscosity dispersion of water $\nu (\omega )$ is related with this maximum. High-frequency oscillations of magnetosomes in the microwave field result in a change of the cytoplasm viscosity in the layer nearby magnetosome's surface. This affects the mean frequency of random rotations of magnetosomes and may cause a biological response.

Grant RFFBR No.04-04-97298.

\end{document}